# Analysis of LLMs vs Human Experts in Requirements Engineering

Cory Hymel
Research
Crowdbotics
Berkeley, CA

Hiroe Johnson
Research
Crowdbotics
Berkeley, CA

*Abstract*—The majority of research around Large Language Models (LLM) application to software development has been on the subject of code generation. There is little literature on LLMs' impact on requirements engineering (RE), which deals with the process of developing and verifying the system requirements. Within RE, there is a subdiscipline of requirements elicitation, which is the practice of discovering and documenting requirements for a system from users, customers, and other stakeholders. In this analysis, we compare LLM's ability to elicit requirements of a software system, as compared to that of a human expert in a time-boxed and prompt-boxed study. We found LLM-generated requirements were evaluated as more aligned (+1.12) than human-generated requirements with a trend of being more complete (+10.2%). Conversely, we found users tended to believe that solutions they perceived as more aligned had been generated by human experts. Furthermore, while LLM-generated documents scored higher and performed at 720× the speed, their cost was, on average, only 0.06% that of a human expert. Overall, these findings indicate that LLMs will play an increasingly important role in requirements engineering by improving requirements definitions, enabling more efficient resource allocation, and reducing overall project timelines.

## I. INTRODUCTION

Recently, software development has been significantly impacted by AI and, more specifically, by large language models (LLMs). These advancements have reshaped how developers approach tasks across the software development lifecycle (SDLC)—from requirements gathering to design, coding, testing, and maintenance. For example, LLMs have been instrumental in automating code generation, debugging, and documentation tasks, demonstrating their potential to streamline traditionally time-intensive processes.

Despite their growing adoption, much of the research and practical applications of LLMs in software engineering have focused on later SDLC phases, such as code development (56.65% of studies), software maintenance (22.71% of studies), and quality assurance (15.14% of studies)—with relatively limited exploration of their role in requirements engineering (RE), which accounts for only 3.90% of studies [1]. This disparity is striking, considering the foundational role of RE in setting the roadmap and ensuring alignment between stakeholder needs and system functionality [2, 3, 4].

Requirements engineering encompasses a systematic approach to eliciting, documenting, and maintaining requirements throughout a software project's lifecycle. Within this discipline, early requirements elicitation—the process of understanding and defining what stakeholders need—is both critical and challenging. Traditional methods such as stakeholder interviews, workshops, and document analysis are time-intensive and rely heavily on human expertise, which can introduce variability and inconsistency.

Given the complexities of early requirements elicitation, this study seeks to evaluate whether LLMs can augment—or even outperform—human experts in generating initial software requirements. By comparing LLM-generated and human-generated requirements under controlled conditions, we aim to provide actionable insights into the capabilities, limitations, and future potential of integrating LLMs into the RE process. Additionally, we hope to highlight the broader implications of leveraging AI in traditionally human-centric tasks—offering insight into how technology may reshape the software development lifecycle in the coming years.

## II. BACKGROUND ON REQUIREMENTS

Requirements engineering (RE) is fundamental to successful software development, encompassing the systematic approach to defining, documenting, and maintaining requirements throughout a project's lifecycle. As Nuseibeh and Easterbrook [5] note, RE ensures that the final product aligns with both user needs and business objectives.

Within RE, elicitation is a crucial first step. Zowghi and Coulin [6] describe it as the process of seeking, uncovering, acquiring, and elaborating requirements for computer-based systems. RE can be broken down into two broad categories: *early requirements* and *late requirements* [7]. Early requirements are those that are created before the software is created, while late requirements are those created once a system is deployed. In this study, we have focused on early requirements—and more specifically, the first generation of requirements created.

Traditionally, early requirements require elicitation from customers and stakeholders by human experts. The experts are typically, but not always, project managers and business analysts. They employ various techniques to elicit needs from customers such as [6, 8]: one-on-one interviews with key stakeholders, group workshops, brainstorming sessions, analysis of existing documentation and systems, prototyping, and user feedback sessions.

Because LLMs are powered by deep learning algorithms and a gigantic training corpus, they offer significant benefits in identifying requirements. Their ability to pull from huge data pools could give them an edge over humans, who can typically only pull from their own limited experience. The speed at which LLMs can generate requirements also offers a significant advantage in the early phases of RE, as traditional methods are often time-consuming and expensive [12].

### III. RESEARCH OBJECTIVES

This study compares the performance of a large language model (LLM) against the performance of human experts in the early stage RE process. Our primary goal is to assess the viability of using LLMs as a tool in the early stage of RE, specifically focusing on their ability to generate meaningful epics and user stories based on limited input. We structured our research around three key questions:

**Q1:** How does the perceived alignment of LLM-generated requirements compare to those produced by human experts?

**Q2:** Can Participants accurately distinguish between LLM-generated and human-generated requirements?

**Q3:** How does familiarity with AI tools influence perceptions of requirement quality and the ability to identify AI-generated content?

To address these questions, we designed a comparative study with the following parameters:

- **Participant Condition**: Participants provided a natural language description in text format of a piece of software they wanted to build.
- **LLM Condition**: A single-shot prompt was given to an advanced LLM, asking it to generate requirements for the specified software system.
- **Human Expert Condition**: An experienced project manager was given 1hr to generate requirements without the use of an LLM based on the same software system.

The outputs from these three conditions were structured as epics and user stories to maintain consistency and reflect common industry practices. To evaluate the results, we employed a mixed-methods approach: quantitative analysis of alignment scores and completeness ratings, quantitative assessment of Participants' ability to identify the source of each document, and exploration of potential correlations between Participants' AI tool experience and their perceptions of the documents.

This research is intended to further the understanding of how LLMs can play a role in the RE process and fill the need for empirical evidence to evaluate the performance of LLMs in a traditionally human-centered task like requirements elicitation. By directly comparing the outputs of an LLM and a human expert under controlled conditions, we provide data on the strengths and limitations of each approach.

### IV. STUDY DESIGN

We crowd-sourced a group of business stakeholders (Participants) who were interested in having a piece of software developed. These Participants submitted their ideas via web form in text format only. Once an idea was submitted, we used the below process (Fig. 1) to create two anonymized requirements documents based on their submission.

1. Document "01": Created by an LLM, ChatGPT-4.0 using a single-shot prompt.
2. Document "02": Created by a human expert during a one-hour, closed session.

Once an idea was submitted, Participants were no longer able to provide input or feedback to their submission. Both documents were structured using epics and user stories to maintain consistency and reflect industry standards. This choice reflects current industry practices and provides a standardized format for comparison. Documents were titled either "Document 01" or "Document 02" and Participants were not told from which source each was generated.

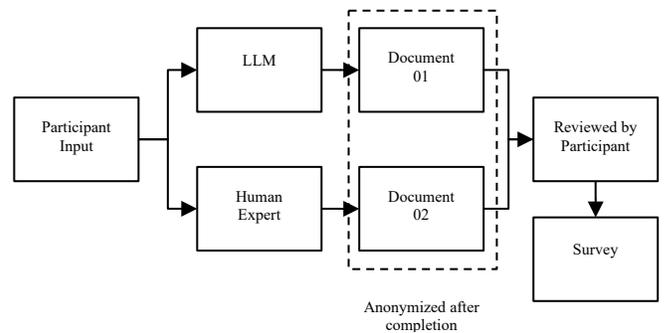

Fig. 1 Study process for documentation creation

#### A. SINGLE-SHOT AND ONE-HOUR

One of the harder aspects of designing this study was creating environmental parity for the LLM and human experts. LLMs can generate huge amounts of information quickly, making it hard to measure performance in a reasonable timeframe against a human expert. We decided on comparing a single-shot prompt to one hour of human-expert working time as a reasonable study baseline.

However, this decision was also a limiting factor for this study. The RE process usually takes place over multiple weeks or months and a large volume of back-and-forth communication between Participants and human experts is common. In the future, this study should be expanded to review the comparative analysis over a complete RE engagement, rather than a simple prompt and time box condition. These limitations are further expanded upon below, preceding the Results section.

B. *SELECTION OF THE LLM AND HUMAN EXPERT*

**LLM Selection**: For this experiment, we used GPT-4 as the LLM Participant. The LLM was provided with a single-shot prompt designed to elicit comprehensive software requirements based on a given project description. The prompt given to GPT-4 is available in Appendix Fig. a-1.

It's important to note that in the prompt, we explicitly state the number of features to create—which differs from what was provided to the human experts. In our early tests using GPT4.0 to generate categories and features, we found that without explicit instruction on a minimum number of features to create, the AI would generate very few items. With further testing, we also found that even when explicitly told to generate a minimum number of features, GPT4.0 would generate between a minimum 15 and maximum 80 features—with an average of 38.16.

**Human Expert Selection**: The human experts were crowdsourced from the popular workplace site Upwork. We sought to use a different expert per Provider submission. The expert was tasked with producing requirements for the same project within a one-hour timeframe and provided a sample spreadsheet consisting of two columns, one for epics and one for features. Human experts were sourced with the following criteria:

- Location: Global
- Talent Type: Freelancers
- Talent Type: Independent
- English level: Conversational
- Skills and Expertise:
    o Requirements Specification
    o Project Management
    o Agile Software Development
    o Product Requirements Document
    o Project Requirements

The job posting can be found in Appendix Fig. a-2. Once the applicable agreements were in place with Upwork, the human expert was given unedited submission from the Participant. If the human expert had any further questions about the project, they were told no additional information was available and that they should use their best guess and expertise. After initial response tests with human experts, we selected a minimum of 20 features. The range count of submitted features varied with an average of 33.07.

C. *PARTICIPANT SELECTION*

In addition to crowdsourcing from the Upwork platform, Participants were also found through the social network site LinkedIn, where a general post was submitted asking anyone with a software idea to submit their concept. Participants sourced through social media were provided with the two requirements documents and a short survey. No monetary compensation was given. Due to a lack of participation, we shifted to source Participants from Upwork, where they were also given the two requirements documents and survey, but also monetary compensation for their time. Providers were also asked how frequently they used AI tools and their self-reported level of AI-tool expertise.

TABLE I. BREAKDOWN OF PARTICIPANTS

| **Frequency of AI tool usage** | | **Self-reported Expertise** | |
| --- | --- | --- | --- |
| Daily | 78% | Novice | 6% |
| Weekly | 14% | Intermediate | 38% |
| Monthly | 4% | Advanced | 46% |
| Never | 4% | Expert | 10% |

Table I. Breakdown of Participants. These tables are not correlated and presented in this way for formatting purposes only.

D. *DATA COLLECTION PROCESS*

Participants were provided with both documents ("01" and "02") without knowing which was created by the LLM and which by the human expert. They were then asked to complete a survey consisting of the following six questions:

1. "How aligned is the document with "01" compared to your initial idea?" (Scale of 1-10, with 10 being most aligned and 1 being least aligned)
2. "How aligned is the document with "02" compared to your initial idea?" (Scale of 1-10, with 10 being most aligned and 1 being least aligned)
3. "Who do you think created each document "01" and "02"?" (Human or AI)
4. "How complete do you think each document is?" (Not Complete, Fairly Complete, Fully Complete)
5. "How often do you use tools like ChatGPT, Claude, Gemini, or other large language models?" (Daily, Weekly, Monthly, Never)
6. "What would you consider your level of "expertise" with tools like ChatGPT, Claude, Gemini, or other?" (Novice, Intermediate, Advanced, Expert)

This survey design allowed us to collect both quantitative and qualitative data on the perceived quality and effectiveness of the LLM-generated requirements, as compared to those created by a human expert.

E. *DATA ANALYSIS*

The survey responses were collected and analyzed to evaluate the following:

**Alignment Scores:** The average alignment scores for each document (01 and 02) were calculated and compared to determine which method (LLM or human) produced requirements more in line with stakeholder expectations.

**Origin Identification:** The accuracy with which Participants identified the origin of each document was analyzed to assess whether stakeholders could reliably distinguish between AI-generated and human-generated requirements. Note: In some instances, due to an unknown bug in the survey, Participants were able to answer both AI and Human. Data for those subjects were removed during direct analysis or cross-analysis.

**Completeness Ratings:** The perceived completeness of each document was analyzed to determine which method produced more comprehensive requirements.

**Impact of AI Familiarity and Expertise:** The correlation between Participants' familiarity and expertise with AI tools and their responses to the survey questions was examined to understand how prior experience with AI might influence their perceptions.

*F. LIMITATIONS*

Our study had several limitations:

1. **Single-instance comparison:** We compared only one LLM-generated document against one human-generated document, which may not be representative of all possible outputs. Expanding to allow for multiple iterations of each document would allow for stronger generalizability.
2. **Subjective nature of evaluations:** Perceptions of alignment and completeness are inherently subjective and may vary based on individual expectations and experiences. Future studies could incorporate more objective metrics or automated tools to evaluate the quality of requirements.
3. **Potential bias in self-reported expertise:** Participants' self-assessment of their AI tool expertise may not accurately reflect their actual proficiency. Including independent validation of Participant expertise could address this limitation.
4. **Small sample size:** The small sample size (n = 50) could influence findings at a larger scale. Broader demographics and a larger Participant pool would provide more robust insights and help verify patterns observed in this study.
5. **Single shot = 1hr:** In order to fit into the experiment budget and resourcing, we settled on the assumption that a single shot prompt would equate to roughly 1 hour of human work. This assumption, while pragmatic, introduces variability in effort and output quality. Alternative configurations or time constraints could enhance comparability in future studies.

These limitations will be considered when interpreting our results and drawing conclusions.

V. RESULTS

*A. ALIGNMENT SCORES:*

The study found a statistically significant (t(48) = 3.179, p = 0.002) understanding of alignment score with the AI-generated document scored higher on average, with a mean difference of 1.122 (Appendix Fig. 3).

FIGURE I: OBJECTIVE ALIGNMENT SCORING

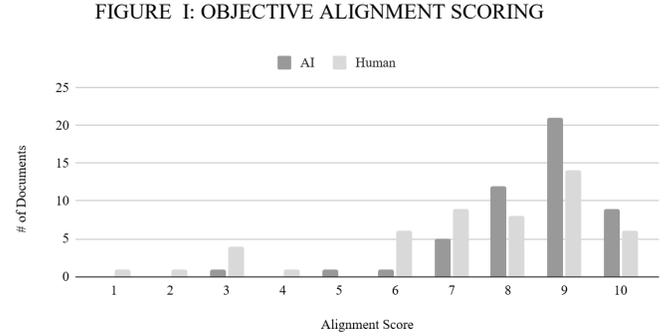

Fig. 1. Alignment scores for both AI and human generated documents.

Notably, the AI document received more consistently high scores, while the human document had a wider range of scores. Diversity in the scoring of the human documents could mean that humans produce more diverse output due to the different interpretations of the project needs and their work experiences—as opposed to shared conglomerate knowledge of an LLM. This finding is consistent with other literature available on the homogenization effects of large language models [13, 14].

FIGURE 2: SUBJECTIVE ALIGNMENT SCORING

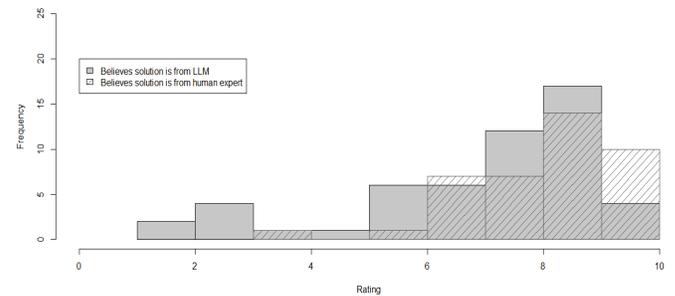

Fig. 2. Heuristics of perceived alignment comparing scores given to solutions identified as AI-generated and human-generated regardless of whether these identifications were correct.

Users tended to believe solutions they perceived as more aligned were generated by human experts. Solutions perceived to be AI-generated received lower scores, with seven ratings of 3 or lower, while solutions perceived to be human-generated had only one score of 3 and none lower. Scores for what providers thought were human-generated solutions were higher (mean: 8.52) than for what they thought were AI-generated solutions (mean: 7.38)—with AI scores showing greater variability (variance: 4.85 vs. 1.96 for human scores). This finding is consistent with other literature showing bias in favor of human output over AI due to multiple reasons, such as the work having less intention and less creativity [17, 18, 19, 20, 21].

B. DOCUMENT SOURCE IDENTIFICATION

Document source identification showed a pattern ($t(41) = 2.44, p = 0.019$) in Participants' ability to distinguish between AI-generated and human-generated requirements documents.

FIGURE 3: DOCUMENT IDENTIFICATION

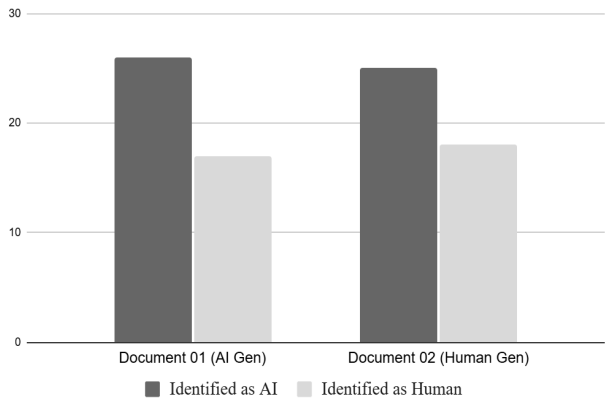

Fig. 3. Distribution of those attributing sources to each document. Full breakdown in Appendix Table A-2

Notably, 60.5% of Participants correctly identified Document 01 as AI-generated, while only 41.9% accurately recognized Document 02 as human-generated. Participants chose AI as the primary document creator both times, suggesting a tendency to over-attribute content to AI sources.

C. COMPLETENESS RATINGS

Completeness ratings were rather interesting with roughly the same scores across both the AI-generated and human-generated documents. We did not find any significant statistical difference between ($t(94) = 0.155, p = 0.877$) with AI scoring marginally higher (+10.2%) in the "fairly complete" category compared to the human-generated documents. Interestingly, the human-generated document received a higher proportion of "Not Complete" ratings (+14.3%) compared to the AI-generated document.

This disparity could be attributed to several factors, but is primarily due to the LLMs ability to generate a broad range of requirements quickly, covering more ground than a human expert constrained by time limitations. However, it is crucial to note that completeness in this context is subjective and based on Participants' perceptions rather than an objective measure of requirements coverage.

FIGURE 4: COMPLETENESS RATING

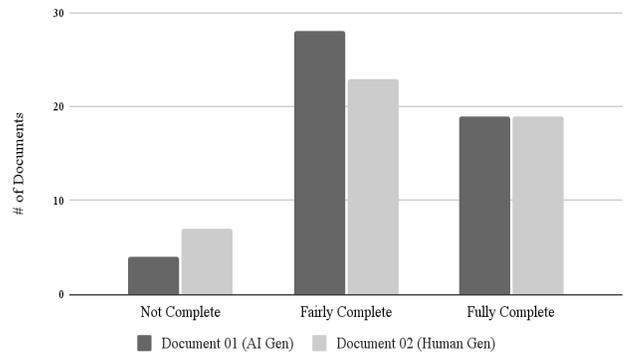

Fig. 4. Number of documents marked as not complete, fairly complete, or fully complete.

There was a significant relationship between Participants' completeness ratings and source attribution for both AI-generated documents ($t(84) = 5.24, p = 1.18E-06$) and human-generated documents ($t(84) = 6.30, p = 1.30E-08$). Of the Participants who rated the AI-generated document as "Fully Complete", 57.1% correctly identified it as AI-generated (Appendix Table A-3)

Conversely, of the Participants who rated the human-generated document as "Fully Complete", only 43.9% correctly identified it as human-generated (Appendix Table A-4). This suggests a potential cognitive bias where high levels of perceived completeness are more readily associated with AI authorship. Such a bias could have significant implications for how professionals evaluate and trust different sources of requirements documentation.

D. IMPACT OF AI FAMILIARITY AND EXPERTISE

One important insight we wanted to investigate was how familiarity with AI tools might influence perceptions of requirement quality and the ability to identify AI-generated content.

We found, when cross-analyzing usage frequency and alignment scoring, daily users tended to rate both documents higher ($t(37) = 3.407, p = 0.002$, Appendix Table A-5), with a preference for the AI-generated document indicating frequent exposure to AI tools may correlate with a higher perceived alignment of AI-generated content. This may possibly be due to increased familiarity with AI output patterns or evolving expectations of document quality. We were not able to get reliable signals from the other usage groups (weekly, monthly, never).

Analyzing expertise and document identification, we also saw intermediate users ($t(33) = 2.076, p = 0.046$) correctly identify the AI-generated document 63.20% of the time while only identifying the human-generated document correctly 31.60%. Contrary to initial expectations, higher self-reported expertise in AI tools did not consistently correlate with better identification accuracy. Due to the limited study size, we are only able to reliably show that Participants with self-reported 'intermediate' AI skills 'had a significantly higher ability to recognize the AI-generated documents than human-generated documents (Appendix Table A-6).

## VI. Discussion

We found that looking at the perceived alignment of LLM- vs. human-generated output, Participants objectively found LLM-generated content to be more aligned. A surprising find, however, was that Participants more often *believed* highly aligned documents to be human generated.

These findings are consistent with the literature, demonstrating that content origination has no impact on trustworthiness or informativeness [22]. The higher distribution of alignment scores for human-generated documents also suggests higher levels of biases due to a single source creator drawing on singular experiences as compared to the more homogenous knowledge of LLMs. While LLMs may be able to produce well informed baseline content, here they show a certain lack of creativity and intuition that is important during the RE process.

Our second question was about Participants' abilities to identify LLM- vs human-generated documents. It was also interesting, in that both documents were over-attributed to AI. AI -generated documents were, on average, five user stories longer — which initially we expected would draw attention to LLM-created documents. Our findings were to the contrary. Other literature has shown that often vague content is attributed to LLM's [23]. Given that human Participants were only given the project description, it is likely that their output was just as vague, leading Participants to over-index on their belief that content was LLM-generated. In longer sessions, where there is more 'back and forth' between Participants and generators (LLMs and humans), we would expect identification to improve—with creativity and domain expertise being top indicators for determining document origin.

We also found that Participants' level of expertise with AI did not correlate to significantly higher identification results. This was also surprising as one would assume that frequent use of AI and exposure to typical LLM output patterns would make identification easier. It did not. There were also not any significant signals in the frequency of use of AI tools that led to stronger identification abilities. These two findings together highlight how LLM-generated content in the requirements engineering process could fit in without significant repercussions from stakeholders or others.

### A. Impact to Requirements Engineering

Our findings in this study highlight several key insights into the potential role of LLMs in requirements engineering (RE). At the initial stages of RE, particularly in requirements elicitation, LLMs demonstrated a significant ability to create aligned and comprehensive requirements quickly.

Today we see LLMs serving as accelerators for requirements generation versus fully replacing human experts. By analyzing artifacts such as meeting notes, audio transcripts, and existing documentation, LLMs will be able to act as "first draft agents." Human experts, leveraging their contextual understanding and domain-specific knowledge, can then validate and refine those drafts into more holistic outputs. We see this hybrid approach not only enhancing efficiency but also introducing new methods for identifying and documenting software.

In the future, we expect LLMs to play a much larger role in the RE process. In this study, we saw LLM-generated requirements were regarded as more aligned (+1.12) and with a trend to being more complete (+10.2%), which when coupled with the cost to generate each document (AI = $0.06, Human = $100, Appendix Table A-7). It's hard to ignore the financial incentive for companies to adopt LLMs and ultimately replace a large portion of the work human experts do today. However, this study is a small sample of the work that goes into the full process of RE.

We expect that humans will still play a critical role in enabling a successful RE process — as LLMs and human experts possess complementary strengths in requirements generation: LLMs excel at producing comprehensive and syntactically correct requirements with human experts bringing domain knowledge, contextual understanding, and the ability to identify nuanced stakeholder needs. As LLMs (and AI) continue to evolve, they will likely take on a more integrated role in the software development lifecycle, with human experts transitioning to roles that focus on orchestration and oversight.

## Acknowledgment

We would like to thank Crowdbotics for sponsoring this research. Darcy Jacobsen from The Wednesday Group for continuous proofing and editing before publication. All the amazing people that participated in the study, thank you.

VII. APPENDIX

|  | Mean | Variance | Standard Deviation |
|---|---|---|---|
| **Document 01 (AI Gen Document)** | 8.450 | 1.838 | 1.344 |
| **Document 02 (Human Gen Document)** | 7.347 | 5.231 | 2.266 |

**Table A-1:** Average values of document alignment with slightly higher alignment of the AI generated documents.

|  | % Correct | % Incorrect |
|---|---|---|
| **Document 01 (AI Gen)** | 60.5 | 39.5 |
| **Document 02 (Hum Gen)** | 41.9 | 58.1 |

**Table A-2:** Percentages that were able to correctly identify origins of documents.

| Level of Completeness | Correctly Identified | Incorrectly Identified |
|---|---|---|
| Not Complete | 75% | 25% |
| Fairly Complete | 60% | 40% |
| Fully Complete | 57.1% | 42.9% |

**Table A-3:** AI Generated Document - Cross analysis of perceived level of completeness and source identification. e.g. Of those who rated the AI generated document as "Fully Complete", 57% correctly identified it as AI-generated

| Level of Completeness | Correctly Identified | Incorrectly Identified |
|---|---|---|
| Not Complete | 33.3% | 66.6% |
| Fairly Complete | 42.9% | 57.1% |
| Fully Complete | 43.6% | 56.3% |

**Table A-4:** Human Generated Document - Cross analysis of perceived level of completeness and source identification. eg. Of those who rated the human generated document as "Fully Complete", 43.6% identified it as human generated.

| Usage | Document 1 | Document 2 | pValue | tStat |
|---|---|---|---|---|
| Daily | 8.64 | 7.31 | 0.002 | 3.407 |
| Weekly | 7.5 | 7.7 | >0.05 | n/a |
| Monthly | 8 | 6 | >0.05 | n/a |
| Never | 9 | 8.5 | >0.05 | n/a |

**Table A-5**: Average alignment score ratings per document based on usage.

| Expertise Level | Correctly Identified Doc 01 | Correctly Identified Doc 02 | pValue | tStat |
|---|---|---|---|---|
| Novice | 50% | 50% | N/A | N/A |
| Intermediate | 63.20% | 31.60% | 0.046 | 2.076 |
| Advanced | 47.80% | 34.80% | 0.507 | 0.671 |
| Expert | 50% | 75% | 0.519 | -0.707 |

**Table A-6**. Identification accuracy based on self-reported AI skill level. Ex. 63.2% of Intermediate users correctly identified Doc 01 as AI generated and 31.6% correctly identified Doc 02 as human generated.

| TI | Avg. Tokens per Prompt (input) | 272 |
|---|---|---|
| TIP | Cost per Token (GPT-4) (input) | $0.00003 |
| TR | Avg. Tokens per Response (output) | 862 |
| TRP | Cost per Token (GPT-4) (output) | $0.00006 |

**Table A-7:** LLM Generated Document = (TI * TIP) + (TR + TRP) = $0.05988

> "You are an expert product manager in the field of software development. Your tasked with writing at a minimum, 40 user stories for a prompt I will give you. You will respond with a table containing two columns: [category] [feature]. You can treat a Category like an epic that has a collection of Features. Here is your first prompt:"

**Fig. A-I:** The single-shot prompt given to ChatGPT-4 with the Participants submission following the end colon.

> You'll be provided a description of a software project that needs to be scoped. We're looking only for Epics and Features (no descriptions, acceptance criteria, etc.) to show to customers human-readable and "makes sense in context" of the project.
>
> Feature - single sentence that describes a certain feature
>
> Epic - category that contains a number of features
>
> Minimum of 20 features
>
> Minimum of 5 epics
>
> The goal is to write as many as possible in 1hr but be creative and thank outside the box!!!
>
> Please spend no longer than 1hr on this task.
>
> Please do not use ChatGPT or AI tools.

**Fig. A-II**: Job post made to crowdsourcing website Upwork.